\documentclass[letterpaper, journal]{IEEEtran}
\usepackage{amsmath, amsfonts}
\usepackage{bm}
\usepackage{array}
\usepackage[caption=false, font=normalsize, labelfont=sf, textfont=sf]{subfig}
\usepackage{textcomp}
\usepackage{stfloats}
\usepackage{url}
\usepackage{verbatim}
\usepackage{graphicx}
\usepackage{booktabs}
\usepackage{multirow}
\usepackage{cite}
\graphicspath{{figs/}}
\begin{document}

\title{Physics-Informed Neural State-Space Modeling of Battery-Electric
Vehicle Dynamics for Closed-Loop Automated Parking Simulation}

\author{Sirong~Pan, Guannan~Tian,
        and~Pan~Song%
\thanks{Manuscript received XX XX, 2026. This work was supported by the
Wuhu Intelligent Logistics Technology Research and Development Center under
Grant WHSYFZX202402. (Corresponding author: Pan Song.)}%
\thanks{S. Pan is with the Modern Logistics and Intelligent Manufacturing
College, Wuhu Vocational Technical University, Wuhu 241003, Anhui, China
(e-mail: 900289@whit.edu.cn).}%
\thanks{G. Tian is with the Department of Vehicle Engineering, Nanjing
University of Aeronautics and Astronautics, Nanjing 210016, China
(e-mail: tianguannan@nuaa.edu.cn).}%
\thanks{P. Song is with the Kaiyang Laboratory, Chery Automobile Co.,
Ltd., Wuhu 241009, Anhui, China (e-mail: songpan14@gmail.com; ORCID:
0000-0003-0814-6824).}%
\thanks{The source code and trained models are available at
\protect\url{https://github.com/pansong/PyNSSM-Parking}, and the parking
simulator at \protect\url{https://github.com/pansong/auto-parking-sim}.}}

\markboth{IEEE Transactions on Intelligent Vehicles}%
{Pan \MakeLowercase{\textit{et al.}}: Physics-Informed Neural State-Space Modeling for Closed-Loop Automated Parking Simulation}

\maketitle

\begin{abstract}
This paper contributes to vehicle dynamics modeling by introducing a
physics-informed neural state-space model tailored for the parking regime
of a production battery-electric sedan, identified entirely from field-test maneuvers. At parking speeds the model captures what the kinematic
idealization omits, including actuator lag, drivetrain creep, brake-hold
transitions through standstill, and frequent reversals of the motion
direction. A gear-conditioned velocity constraint is imposed during
training, and the yaw rate is read out as a learned residual on a
kinematic-bicycle prior, so that the network devotes its capacity to the
deviation from physics rather than to its reproduction. These
training-time physics make the customary inference-time state limiter
redundant. The commanded-to-actual behavior of the drive, brake, and
steering actuators is reproduced by dedicated submodels, for which signal
fidelity proves an unreliable proxy for closed-loop value; tuning the
brake on its velocity consequence rather than on its own signal reverses
the verdict reached at the signal level. The model generalizes to held-out maneuvers in fully open-loop
simulation, and, despite being identified from only 16 field tests,
the assembled command-to-vehicle chain earns Good ratings on the vehicle
states under the ISO/TS 18571 objective rating metric. Embedded as the real-time plant of an interactive
simulator, it enables a production-representative planning stack to park
the vehicle through the learned dynamics. This makes the model suitable
for pre-calibrating an automated-parking planning and control stack in the
virtual development phase without the manufacturer's proprietary chassis
and actuator parameters.
\end{abstract}


\section{Introduction}
Accurate vehicle dynamics modeling is fundamental to the development,
tuning, and evaluation of automated driving functions, including
automated parking. The parking regime, however, is poorly served by
established modeling practice. At speeds below walking pace the classical
dynamics models collapse to kinematics, yet the real behavior is dominated
by the effects those kinematics discard: actuator lag and hysteresis,
the creep torque of an electric driveline, brake-hold transitions through
standstill, and frequent reversals of the motion direction. Production
practice nonetheless adopts the kinematic-bicycle model for
planning, where it is appropriate, and for execution-level
simulation, where it idealizes away the phenomena that decide whether a
maneuver lands in the bay.

Data-driven modeling avoids this idealization. Neural networks learn
vehicle dynamics from real-vehicle measurements
\cite{oh1999,spielberg2019,nie2022,song2024pynssm}, neural state-space and
continuous-time formulations cast system identification as the learning of
compact state and output maps \cite{chen2018node,masti2021,forgione2021},
and physics-informed or hybrid models constrain the learned dynamics
toward physical plausibility
\cite{raissi2019pinn,chrosniak2024deep,baier2022}. Across this literature,
however, the model is almost always judged by open-loop prediction
accuracy on forward-driving data, and almost never in the role it is built
to fill, as the executed plant inside a closed planning and control loop.
A model accurate one step ahead can still
fail as a closed-loop plant, where errors compound, the controller reacts
to the model's own outputs, and evaluation must run in real time. For a
parking plant, on which a production planner is tuned and validated, this
closed-loop fidelity, not the open-loop residual, determines whether the
model is usable.

Two further gaps are specific to parking. Its dynamics involve signed
velocities and gear-conditioned transitions among reverse, drive, and
standstill that forward-driving formulations do not represent, and the
commanded-to-actual transfer of the production actuators, which dominates
the response at parking speeds, is rarely modeled. The precursor to this
work \cite{song2024pynssm} introduced a neural state-space (NSS) model of
forward-driving dynamics and identified parking as the open extension.
Automated-parking motion planning is itself mature
\cite{dolgov2010,lian2023}, but it is developed against the kinematic
idealization rather than a high-fidelity learned plant of the target vehicle.

These gaps weigh most on the teams that develop automated-parking
software, which must calibrate planning and control against vehicle
dynamics that the manufacturer seldom releases in parametric form and that
are costly to characterize on a prototype. This paper develops, analyzes,
and deploys a physics-informed NSS model of the parking-regime dynamics of
a production battery-electric sedan, identified entirely from 16 field-test maneuvers and validated in the role it is meant to serve, as the
real-time plant inside an interactive parking simulator through which a
production-representative planning stack parks the vehicle. The
contributions are fourfold:
\begin{enumerate}
\item a parking-capable NSS formulation combining a gear-conditioned,
three-branch velocity constraint with a physics-residual yaw-rate readout,
whose training-time physics render the customary inference-time state
limiter redundant;
\item an actuator modeling study across five architectures and three
state-input variants per channel, showing that open-loop signal fidelity
does not predict closed-loop value, that state dependence differs across
channels, and that consequence-tuning the selected submodel, by
backpropagation through the frozen vehicle model, reverses the verdict
against long-memory architectures;
\item a closed-loop deployment in which a production-representative
planning stack parks the vehicle through the learned dynamics over a
deterministic 36-cell scenario grid and a 252-cell held-out robustness
grid, with the controller-induced clearance trade reported;
\item a small-data experimental methodology comprising stratified
cross-validation, a leave-$r$-repeats-out learning curve, seed-noise
decomposition, and sensor-blind-zone-masked metrics, by which each design
choice is either supported by a controlled comparison exceeding the
measured noise floor or resolved by simplicity.
\end{enumerate}

The rest of this paper is organized as follows. Section~\ref{sec:model}
presents the physics-informed NSS model.
Section~\ref{sec:actuators} develops the per-channel actuator submodels and
their physical envelopes. Section~\ref{sec:setup} details the field-test
data, the training and evaluation protocol, and the simulator
deployment. Section~\ref{sec:results} reports the open-loop validation,
the limiter comparison, the design ablations, the actuator study, the
standardized fidelity assessment, and the closed-loop evaluation. Section~\ref{sec:conclusion} concludes.

\section{Physics-Informed Neural State-Space Model}
\label{sec:model}

\begin{figure}[!t]
\centering
\includegraphics{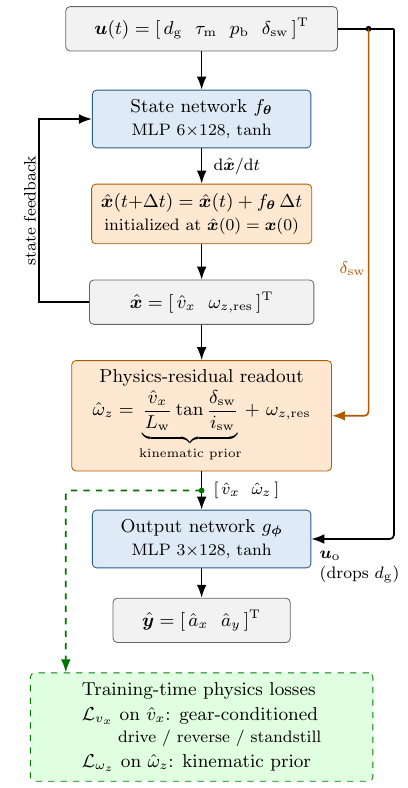}
\caption{Architecture of the physics-informed neural state-space model, with
state $\hat{\bm{x}} = [\hat{v}_x\ \ \omega_{z, \mathrm{res}}]^{\mathrm{T}}$ and
yaw-rate readout
$\hat{\omega}_z = \omega_{z, \mathrm{ref}} + \omega_{z, \mathrm{res}}$.
Blue blocks denote learned networks, orange blocks deterministic
operations, gray blocks signals, and the dashed green band the
physics-informed losses, which act at training time only.}
\label{fig:arch}
\end{figure}

\subsection{Neural State-Space Formulation}

Following the reduced-order strategy established in \cite{song2024pynssm},
the parking-regime dynamics of the test vehicle are represented within a
neural state-space framework. The state vector
$\bm{x} = [v_x\ \ \omega_z]^{\mathrm{T}}$ comprises the longitudinal
velocity and the yaw rate, and the output vector
$\bm{y} = [a_x\ \ a_y]^{\mathrm{T}}$ collects the longitudinal and lateral
accelerations measured by the onboard inertial sensor. The input vector
\begin{equation}
\bm{u} = [\,d_\mathrm{g} \ \ \tau_\mathrm{m} \ \ p_\mathrm{b} \ \ \delta_\mathrm{sw}\,]^{\mathrm{T}}
\label{eq:input}
\end{equation}
consists of the gear state $d_\mathrm{g} \in \{-1, 0, +1\}$ for reverse,
neutral/park, and drive, respectively; the motor torque $\tau_\mathrm{m}$;
the brake master-cylinder pressure $p_\mathrm{b}$; and the steering-wheel
angle $\delta_\mathrm{sw}$. The zero level merges neutral and park deliberately.
The recorded gear signal encodes both identically, and the two are
dynamically equivalent within the modeled states: neither transmits
tractive torque, and both occur only at standstill, where the parking
pawl's mechanical lock acts entirely inside the wheel-speed sensor blind
zone. Their distinct deployment semantics are restored by the command
interface of Section~\ref{sec:setup:sim}. Note that these are the actual actuator
responses recorded on the vehicle bus rather than the controller commands;
the commanded-to-actual transfer is modeled separately by the per-channel
actuator submodels of Section~\ref{sec:actuators}, which permits the two model
families to be identified and validated independently before being chained
in closed loop.

The NSS model comprises two principal components, a state network
$f_{\bm\theta}$ and an output network $g_{\bm\phi}$, each structured as a
fully connected multi-layer perceptron (MLP) with hyperbolic-tangent activations. At each time step, the
state network computes the time derivative of the state rather than the
next state directly; this derivative is then integrated by the forward-Euler
rule at the field-test sampling interval $\Delta t = 10$\,ms:
\begin{subequations}
\begin{align}
\hat{\bm{x}}(0) &= \bm{x}(0), \label{eq:nss:ic}\\
\hat{\bm{x}}(t + \Delta t) &= \hat{\bm{x}}(t)
  + f_{\bm\theta}\!\left(\hat{\bm{x}}(t), \, \bm{u}(t)\right)\Delta t,
  \label{eq:nss:state}\\
\hat{\bm{y}}(t) &= g_{\bm\phi}\!\left(\hat{\bm{x}}(t), \,
  \bm{u}_\mathrm{o}(t)\right), \label{eq:nss:output}
\end{align}
\end{subequations}
where $f_{\bm\theta}$ and $g_{\bm\phi}$ denote the state and output networks
with trainable parameters $\bm\theta$ and $\bm\phi$, respectively;
$\hat{\bm{x}}(t)$ and $\hat{\bm{y}}(t)$ are the predicted state and output at
time $t$; the initial prediction $\hat{\bm{x}}(0)$ is set to the true initial
state $\bm{x}(0)$; $\Delta t$ is the integration step size; and the
output-branch input
$\bm{u}_\mathrm{o} = [\,\tau_\mathrm{m} \ \ p_\mathrm{b} \ \ \delta_\mathrm{sw}\,]^{\mathrm{T}}$
excludes the gear state, whose influence is already encapsulated in the
evolution of $\hat{\bm{x}}$. All subsequent predictions follow from
\eqref{eq:nss:state} in a purely open-loop manner, with no measured state
fed back during inference, so that every reported metric reflects
multi-step rollout accuracy over the complete maneuver. The state network
employs six hidden layers of 128 units and the output network three; the
selection of this capacity is revisited empirically in
Section~\ref{sec:results}.

All channels are scaled to $[-1, 1]$ by min--max normalization computed on
the training set only. In contrast to \cite{song2024pynssm}, each channel is
first clipped to its 0.5th--99.5th percentile range, which prevents rare
actuator transients from compressing the effective resolution of the
remaining samples while keeping the scaled signals bounded and therefore
compatible with the saturating activations. The data-fidelity losses are
mean absolute errors on the scaled states and outputs; the velocity
component of the state loss is masked wherever the measured speed falls
inside the wheel-speed sensor blind zone of Section~\ref{sec:setup},
$|v_x| < 0.5$\,km/h, within which the reference signal carries no usable
information.

\subsection{Physics-Residual Yaw-Rate Readout}

At parking speeds, tire slip is negligible and the yaw rate is governed
almost entirely by kinematics: on the field-test data, the
kinematic-bicycle relation
\begin{equation}
\omega_{z, \mathrm{ref}} = \frac{v_x}{L_\mathrm{w}}
  \tan\!\left(\frac{\delta_\mathrm{sw}}{i_\mathrm{sw}}\right)
\label{eq:bicycle}
\end{equation}
explains the measured yaw rate with a coefficient of determination of
0.997, where $\omega_{z, \mathrm{ref}}$ denotes the kinematic reference yaw
rate; $L_\mathrm{w} = 3.0$\,m the wheelbase; and $i_\mathrm{sw} = 12.1$ the
steering ratio. Such a strong prior is exploited structurally rather than
discarded: the state network integrates the residual yaw rate
$\omega_{z, \mathrm{res}}$ as its second state, so the internal state of
\eqref{eq:nss:state} becomes
$\hat{\bm{x}} = [\,\hat{v}_x \ \ \omega_{z, \mathrm{res}}\,]^{\mathrm{T}}$,
and the observed yaw rate is reconstructed at readout as
\begin{equation}
\hat{\omega}_z(t) = \omega_{z, \mathrm{ref}}\!\left(\hat{v}_x(t),
  \delta_\mathrm{sw}(t)\right) + \omega_{z, \mathrm{res}}(t),
\label{eq:residual}
\end{equation}
with the initial residual decomposed consistently from \eqref{eq:nss:ic} as
$\omega_{z, \mathrm{res}}(0) = \omega_z(0) - \omega_{z, \mathrm{ref}}(0)$. The
reconstruction \eqref{eq:residual} is applied wherever the yaw rate is
consumed, namely the output network input, the physics-informed losses, and
all evaluation metrics; in particular, the output network in
\eqref{eq:nss:output} receives the reconstructed pair
$[\,\hat{v}_x \ \ \hat{\omega}_z\,]^{\mathrm{T}}$ rather than the internal
state, as depicted in Fig.~\ref{fig:arch}. The network capacity is thereby
devoted exclusively to the small slip- and compliance-induced deviation
from the kinematic ideal.
This structural prior reduces the validation velocity error by
approximately 27\% relative to an otherwise identical model that predicts
the yaw rate directly, an ablation reported in
Section~\ref{sec:results:ablations}.

\subsection{Gear-Conditioned Physics-Informed Losses}
\label{sec:model:pinn}

The parent model \cite{song2024pynssm} addresses forward driving only.
Parking maneuvers, in contrast, are dominated by direction changes: the
velocity is negative in reverse, positive in drive, and pinned at
standstill between the two. The physical direction constraint is therefore
imposed during training as a soft, gear-conditioned loss with three
mutually exclusive branches:
\begin{equation}
\begin{aligned}
\mathcal{L}_{v_x} = \mathbb{E}\Big[
 &\mathbb{1}(d_\mathrm{g} > 0.5)\,\max\!\big(0, -\hat{v}_x\big) \\
 &+ \mathbb{1}(d_\mathrm{g} < -0.5)\,\max\!\big(0, \hat{v}_x\big) \\
 &+ \mathbb{1}(|d_\mathrm{g}| \le 0.5)\,\big|\hat{v}_x\big|\Big],
\end{aligned}
\label{eq:pinnv}
\end{equation}
where $\mathbb{E}[\cdot]$ denotes the expectation over the training samples,
$\mathbb{1}(\cdot)$ the indicator function, and all quantities are evaluated
in the scaled space. The drive and reverse
branches penalize only direction violations, leaving the velocity
magnitude unconstrained, whereas the zero-gear branch anchors the velocity
toward standstill, reflecting that the test vehicle occupies neutral or
park only at rest. A second soft loss attracts the yaw-rate prediction toward the
kinematic prior,
\begin{equation}
\mathcal{L}_{\omega_z} = \mathbb{E}\left[\,\big|\hat{\omega}_z
 - \omega_{z, \mathrm{ref}}\big|\,\right],
\label{eq:pinnyaw}
\end{equation}
and the total training objective combines the masked data losses with the
two physics terms,
\begin{equation}
\mathcal{L} = \mathcal{L}_{\bm{x}}^\mathrm{masked} + \mathcal{L}_{\bm{y}}
 + \lambda_{v_x}\,\mathcal{L}_{v_x}
 + \lambda_{\omega_z}\,\mathcal{L}_{\omega_z},
\label{eq:loss}
\end{equation}
where $\mathcal{L}_{\bm{x}}^\mathrm{masked}$ and $\mathcal{L}_{\bm{y}}$ are
the masked state and output data-fidelity losses, and $\lambda_{v_x}$ and
$\lambda_{\omega_z}$ weight the velocity and yaw-rate physics terms. The
weights $\lambda_{v_x} = 0.1$ and $\lambda_{\omega_z} = 0.5$ are fixed by
physical reasoning and simplicity rather than by tuning: in the
hyperparameter sweep their effect was weak relative to the seed-noise
floor, so, following the selection discipline of
Section~\ref{sec:setup:training}, the heavier yaw weight reflects the high
reliability of the kinematic prior it enforces, while the lighter velocity
weight imposes only a soft direction-and-standstill constraint. The question of whether a hard state limiter retains any value
at inference time is then settled empirically in
Section~\ref{sec:results:limiter}.

\section{Per-Channel Actuator Models}
\label{sec:actuators}

\begin{figure*}[!t]
\centering
\includegraphics[width=\textwidth]{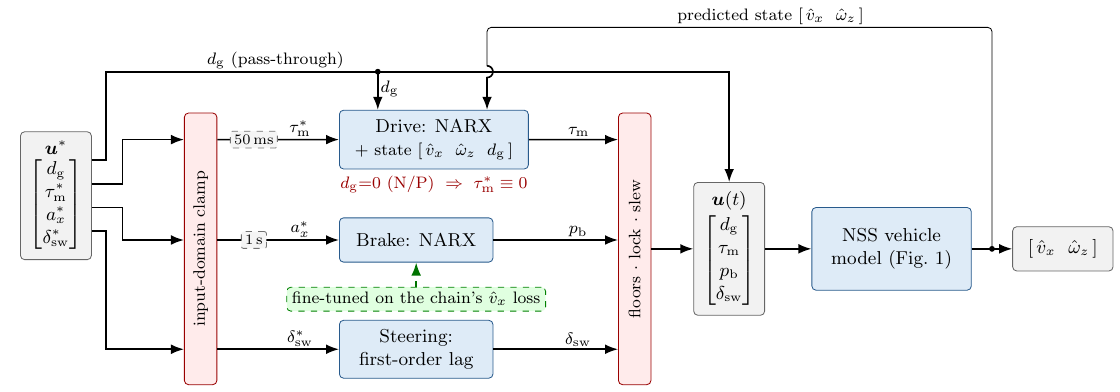}
\caption{Command-to-vehicle chain in deployed configuration.
Controller commands pass the input-domain clamp, the per-channel actuator submodels, and the physical output envelope to form the actual input vector
$\bm{u}(t)$ of the NSS vehicle model. Dashed buffers mark the command histories the
drive and brake submodels consume; the gear state passes through unclamped and also
feeds the drive submodel. The green dashed element acts at training time
only, fine-tuning the brake submodel on the chain's velocity loss,
Section~\ref{sec:results:actuators}. Asterisks denote commanded quantities;
blue blocks denote identified models, red physical envelopes and hard
interface rules, and gray signals.}
\label{fig:chain}
\end{figure*}

\begin{figure}[!t]
\centering
\includegraphics[width=\columnwidth]{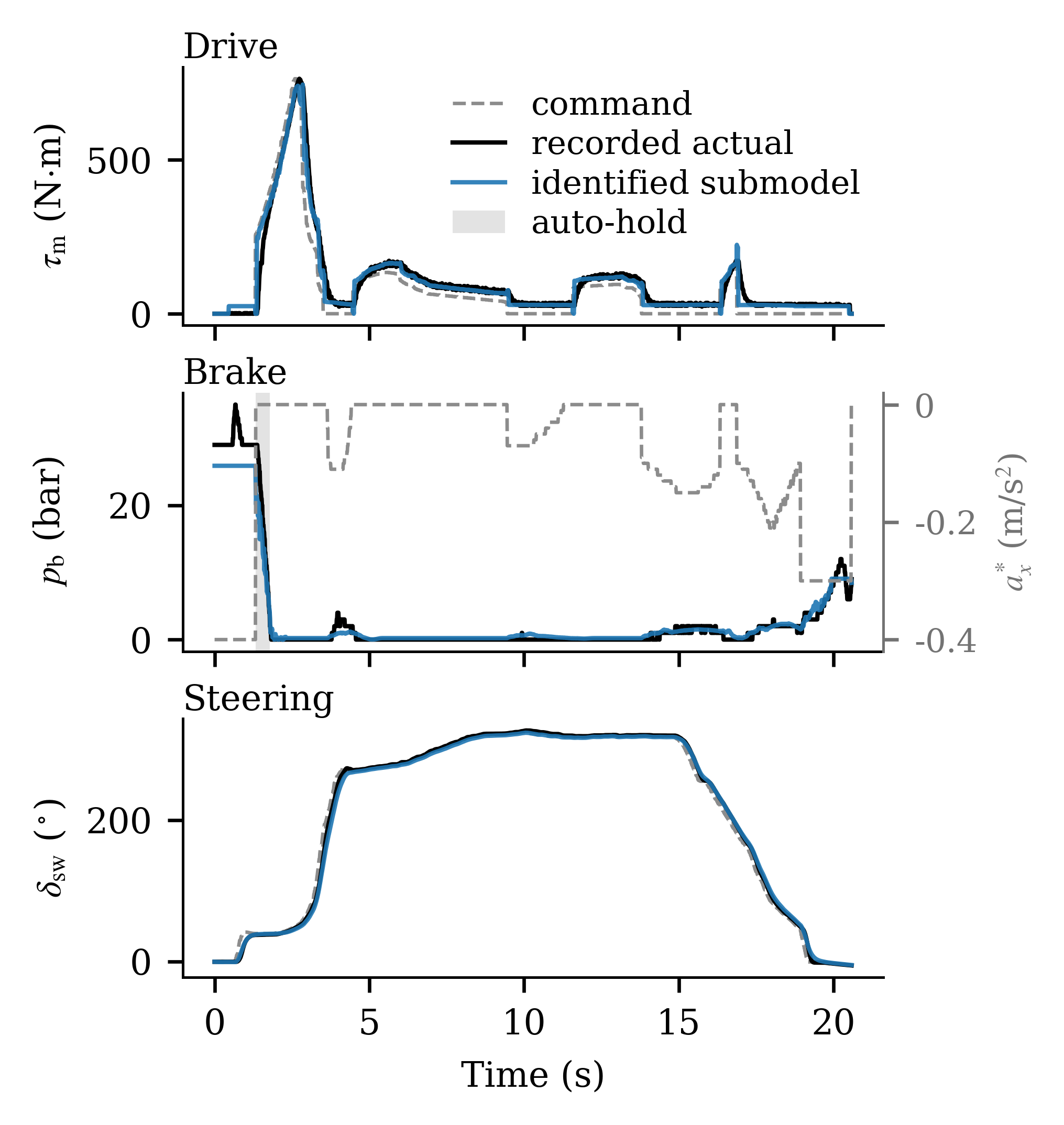}
\caption{Commanded-to-actual transfer of the deployed actuator channels
over a complete held-out park-out maneuver, showing the commanded signal,
the recorded actual response, and the identified submodel of each channel
replayed open loop. The shaded region marks the auto-hold pressure
released at launch.}
\label{fig:replay}
\end{figure}

The NSS model of Section~\ref{sec:model} consumes the actual
actuator responses, whereas a parking controller issues commands. In
\cite{song2024pynssm}, the two were treated as identical, which is
acceptable for offline replay but precludes closed-loop simulation: under a
controller in the loop, the lag, gain, and saturation of each actuator
shape the vehicle response and the controller's reaction to it. The
field-test vehicle records both sides of this transfer on its bus, which
enables the commanded-to-actual dynamics to be identified explicitly and
chained in front of the NSS model; Fig.~\ref{fig:chain} depicts the
deployed chain. The decomposition also carries a practical commitment
beyond fidelity: each channel can be re-identified in isolation when its
subsystem is recalibrated, a property examined together with the
monolithic alternative in Section~\ref{sec:results:actuators}.

\subsection{Channels and Candidate Architectures}

Three actuator channels are modeled, supplying the components
$\tau_\mathrm{m}$, $p_\mathrm{b}$, and $\delta_\mathrm{sw}$ of the NSS input
vector~\eqref{eq:input}, with commanded quantities distinguished from the
recorded actuals by an asterisk. The drive channel
maps the torque command to the actual motor torque,
$\tau_\mathrm{m}^{*} \rightarrow \tau_\mathrm{m}$. The
brake channel maps the deceleration command directly to the
master-cylinder pressure, $a_x^{*} \rightarrow p_\mathrm{b}$; a cascade
through the intermediate booster rod stroke was examined but not adopted,
the rod stroke proving weakly informative about the resulting pressure.
The steering channel maps the steering-wheel angle request to the actual
steering-wheel angle,
$\delta_\mathrm{sw}^{*} \rightarrow \delta_\mathrm{sw}$. Figure~\ref{fig:replay} illustrates each commanded-to-actual
transfer on a held-out maneuver, together with its reproduction by the
per-channel candidates ultimately selected in
Section~\ref{sec:results:actuators}, replayed through the deployed
physical envelopes. Residual mismatches remain visible on the drive channel, a small standstill
offset and a release tail, but their closed-loop cost is bounded by the
end-to-end accounting of Section~\ref{sec:results:actuators}.

One property of the brake channel shapes its modeling. Beyond the
commanded path, the recorded pressure carries an autonomous auto-hold
component that no command explains. Decided by unobserved chassis logic,
these holds appear in Fig.~\ref{fig:replay} as the shaded pressure at
launch, and no command-driven candidate reproduces them, but the omission
is tolerable, the deployed simulator regenerating hold-class pressure by
commanding deceleration at every stop (Section~\ref{sec:setup:sim}). Their
slow release nonetheless exceeds the 50\,ms window of the
identification-stage submodel, the reason the deployed brake carries a
1\,s window, with long-memory architectures, a long short-term memory
(LSTM)~\cite{hochreiter1997} network and a longer-window nonlinear
autoregressive with exogenous inputs (NARX)~\cite{sjoberg1995} model,
evaluated in Section~\ref{sec:results:actuators}.

For every channel, five candidate architectures spanning three orders of
magnitude in parameter count are fitted, with $u^{*}$ and $y$ denoting the
channel's scalar command and actual response, respectively:
\begin{enumerate}
\item a first-order lag, $\tau_1 \dot{y} = u^{*} - y$, with a single
learnable time constant;
\item a second-order lag parameterized by $(\tau_2, \zeta, K)$, capturing
inertia and damping of hydraulic components;
\item a Hammerstein model, in which a small static nonlinearity
$\psi(u^{*})$, a $1{\times}8{\times}8{\times}1$ perceptron, precedes a
first-order lag, accommodating dead zones and saturation;
\item a NARX perceptron operating on the command and five lagged samples,
50\,ms of history;
\item a gated recurrent unit (GRU)~\cite{cho2014} with a 16-dimensional hidden state.
\end{enumerate}
Each candidate is additionally fitted in three input variants: command
history only; command history plus physically motivated state features,
such as vehicle speed and gear for the drive channel; and command history
plus a uniform state interface $[\,v_x \ \ \omega_z \ \ d_\mathrm{g}\,]$
shared by all channels. Whether an actuator submodel benefits from vehicle-state inputs is
thereby treated as an experimental question rather than a design
assumption, and Section~\ref{sec:results:actuators} shows the answer
differs across channels.

\subsection{Physical Input--Output Envelopes}
\label{sec:actuators:limits}

In closed-loop operation a controller may issue commands outside the
domain spanned by the field tests, where a learned model extrapolates
without physical grounding. All deployed submodels are therefore wrapped
in physical envelopes derived from the data and the vehicle specification:
commands are clamped to the identified input domains, for instance the
deceleration command $a_x^{*}$ to $[-0.8, 0]$\,m/s$^2$, while outputs
observe the physical floors, with torque and pressure non-negative, the
steering lock of $\pm 400^\circ$, and a steering rate limited just above
the highest slew rate observed in the field tests so the guard acts only
on out-of-domain requests. A stress study in
Section~\ref{sec:results:actuators} examines which envelope regulates the
chain. The recorded actuals are quantized to 2\,N$\cdot$m and 1\,bar,
setting a noise floor below which fidelity differences are not meaningful.

\subsection{Two-Phase Selection With Closed-Loop Fine-Tuning}

Candidate selection proceeds in two phases. Phase~A scores every
channel--architecture--variant combination against the recorded actuals on
held-out trials, by mean squared error (MSE), mean absolute error (MAE),
and step-response metrics, and shortlists per channel. Phase~B substitutes
each shortlisted candidate into the complete chain, from controller
commands through the submodel into the NSS rollout, and selects per-channel
winners by the resulting vehicle-state accuracy; because submodel errors
interact through the vehicle state, the assembled chain with all submodels
active is the final authority. The protocol closes with an optional
consequence-level stage that fine-tunes a selected submodel by
backpropagating the chain's vehicle-state loss through the frozen vehicle
model into its weights, weighing each actuator error by its closed-loop
effect rather than its signal proximity. A state-aware winner, fitted with
measured states but deployed with NSS-predicted ones, is additionally
fine-tuned for a short schedule on streams in which the frozen NSS supplies
the state features, closing the identification-to-deployment gap without
altering the NSS.

\section{Experimental Setup}
\label{sec:setup}

\subsection{Test Vehicle and Field Test Data}

The test vehicle is a battery-electric sedan, the EXEED
Sterra~ES, equipped with a production automated parking system. Its
geometry enters the model as fixed constants: wheelbase
$L_\mathrm{w} = 3.0$\,m, track width 1.695\,m, steering-wheel lock
$\pm 400^\circ$, and steering ratio $i_\mathrm{sw} = 12.1$. The rack is
mildly nonlinear, about 13 near center and 12.1 toward the lock, and the
adopted value of 12.1 is
corroborated by an independent least-squares fit of \eqref{eq:bicycle} to
the field-test data, which yields 12.0.

A total of 16 parking trials were recorded at 100\,Hz over the in-vehicle
network, organized as four maneuver classes with four repetitions each:
multi-segment park-in maneuvers containing reverse driving and gear
reversals, and single-segment park-out maneuvers, each toward the left and
the right. Trial lengths range from roughly 21 to 64\,s. Every trial
captures the complete command--response pairs of
Section~\ref{sec:actuators} alongside the vehicle states and accelerations,
which enables the actuator submodels and the vehicle model to be identified
from the same data. One sensor property shapes the entire evaluation
methodology. The wheel-speed-derived velocity cannot resolve magnitudes
below approximately 0.5\,km/h, inside which the recorded signal is
quantization noise rather than measurement. Parking maneuvers spend a
substantial fraction of their duration at or near standstill, so naive
error metrics would largely score the reproduction of sensor noise; all
velocity losses and headline metrics in this paper are therefore masked to
the moving region $|v_x| \ge 0.5$\,km/h.

\subsection{Training and Evaluation Protocol}
\label{sec:setup:training}

\subsubsection{Training Configuration}
\begin{figure}[!t]
\centering
\includegraphics[width=\columnwidth]{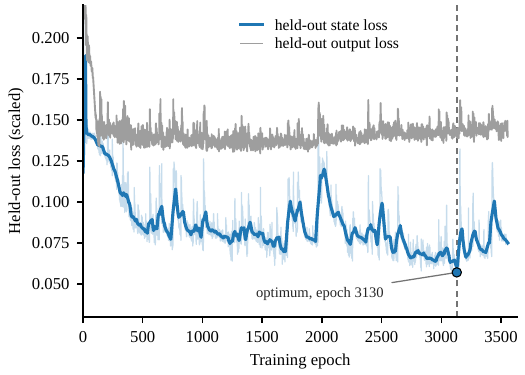}
\caption{Convergence of the held-out losses on one cross-validation
fold. The blue state loss, shown raw and smoothed, bottoms at the marked
optimum, epoch 3130; the gray output loss plateaus. Deployment fits are
capped at 3200 epochs.}
\label{fig:convergence}
\end{figure}

Each trial is tiled into end-aligned windows of 2121 samples, the shortest
trial length, yielding 29 training sequences that jointly cover approach,
transition, and standstill phases. The networks are trained for the full
windows by backpropagation through the open-loop rollout
\eqref{eq:nss:state}, with the Adam optimizer under a cosine
learning-rate schedule starting at $2 \times 10^{-3}$, mean-absolute-error
data losses, and the physics terms of \eqref{eq:loss}. All models are
implemented in PyTorch and trained on a single NVIDIA RTX~A4500 GPU under
the Windows Subsystem for Linux (WSL2). A convergence probe
(Fig.~\ref{fig:convergence}) places the validation optimum near 3000
epochs, beyond which the held-out error stagnates or mildly degrades, so
deployment fits are capped at 3200 epochs with continuous best-checkpoint
tracking on a smoothed validation criterion. The
normalization, loss criterion, learning-rate schedule, training-window
length, and integration scheme were each selected by a dedicated controlled
comparison rather than by convention, and are reported as ablations in
Section~\ref{sec:results:ablations}.

\subsubsection{Data Splits}
Three complementary splits are employed. First, a stratified four-fold
cross-validation holds out one trial per maneuver class per fold, 12
training and 4 validation trials each, so that every reported
generalization figure is an average over class-balanced held-out
maneuvers. Second, a leave-$r$-repeats-out grid with $r = 1, 2, 3$, four folds
each, at a matched 3000-epoch budget, traces the learning curve from 4 to
12 training trials under a consistent protocol. Third, the deployment
model is fitted on all 16 trials with the cross-validation statistics
serving as its unbiased accuracy estimate.

\subsubsection{Evaluation Discipline}
All evaluations roll the model open loop from the true initial state over
each validation trial's complete, untruncated duration, up to
three times longer than the training windows, and report errors in
physical units. Errors are decomposed into the moving region, the reverse
subset, and within- versus beyond-training-horizon segments. Model
selection never relies on any inference-time state limiter; the comparison
of limiters is itself a result, reported in Section~\ref{sec:results:limiter}.
Finally, run-to-run seed variability is treated as a noise floor to be
measured, not a lottery to be exploited. Repeated trainings established a
seed-induced coefficient of variation of approximately 2\% on the
moving-region velocity error and 3--4\% on the yaw-rate error, and design
decisions are accepted only when their effect exceeds this floor;
candidate selection is otherwise resolved by simplicity or physical
reasoning.

\subsection{Parking-Simulator Deployment}
\label{sec:setup:sim}

The closed-loop evaluation embeds the learned models in an interactive
parking simulator driven by a production-representative planning stack: a
Hybrid~A$^{*}$ lattice search~\cite{dolgov2010} with Reeds--Shepp
expansion~\cite{reeds1990}, an optimization-based trajectory
refinement~\cite{zhang2018hobca}, and a predictive steering tracker,
detailed in the released simulator. Two principles govern the integration.
First, the planner keeps its internal kinematic-bicycle
model~\cite{polack2017}, as in a real vehicle; only the executed dynamics
are exchanged for the learned plant, and closed-loop tracking absorbs the
difference. Second, a command interface translates the tracker's speed and
steering commands into the production command set
$\{d_\mathrm{g}, \tau_\mathrm{m}^{*}, a_x^{*}, \delta_\mathrm{sw}^{*}\}$
through a gear state machine and a longitudinal controller. The gear logic
follows the field-test evidence that all 72 recorded direction changes
shift reverse-to-drive at standstill without passing through neutral; park
pins the states at zero for the parking pawl, while neutral forces
$\tau_\mathrm{m}^{*} \equiv 0$ (Fig.~\ref{fig:chain}). The longitudinal
controller adds a brake-hold that raises the commanded deceleration to
standstill, necessary because the learned plant reproduces the driveline
creep torque that defeats a fixed gentle brake. Interface gains are tuned
by coordinate descent on a 12-cell core; all robustness grids in
Section~\ref{sec:results:closedloop} use held-out scenarios never seen
during tuning.

The learned plant runs in the browser as a numerical port of the reference
implementation, verified equivalent over recorded, closed-loop, and
randomized command streams to floating-point precision, with each NARX
submodel fed its true command history through a per-maneuver ring buffer.

\section{Results and Discussion}
\label{sec:results}

\subsection{Open-Loop Validation}
\begin{figure}[!t]
\centering
\includegraphics[width=\columnwidth]{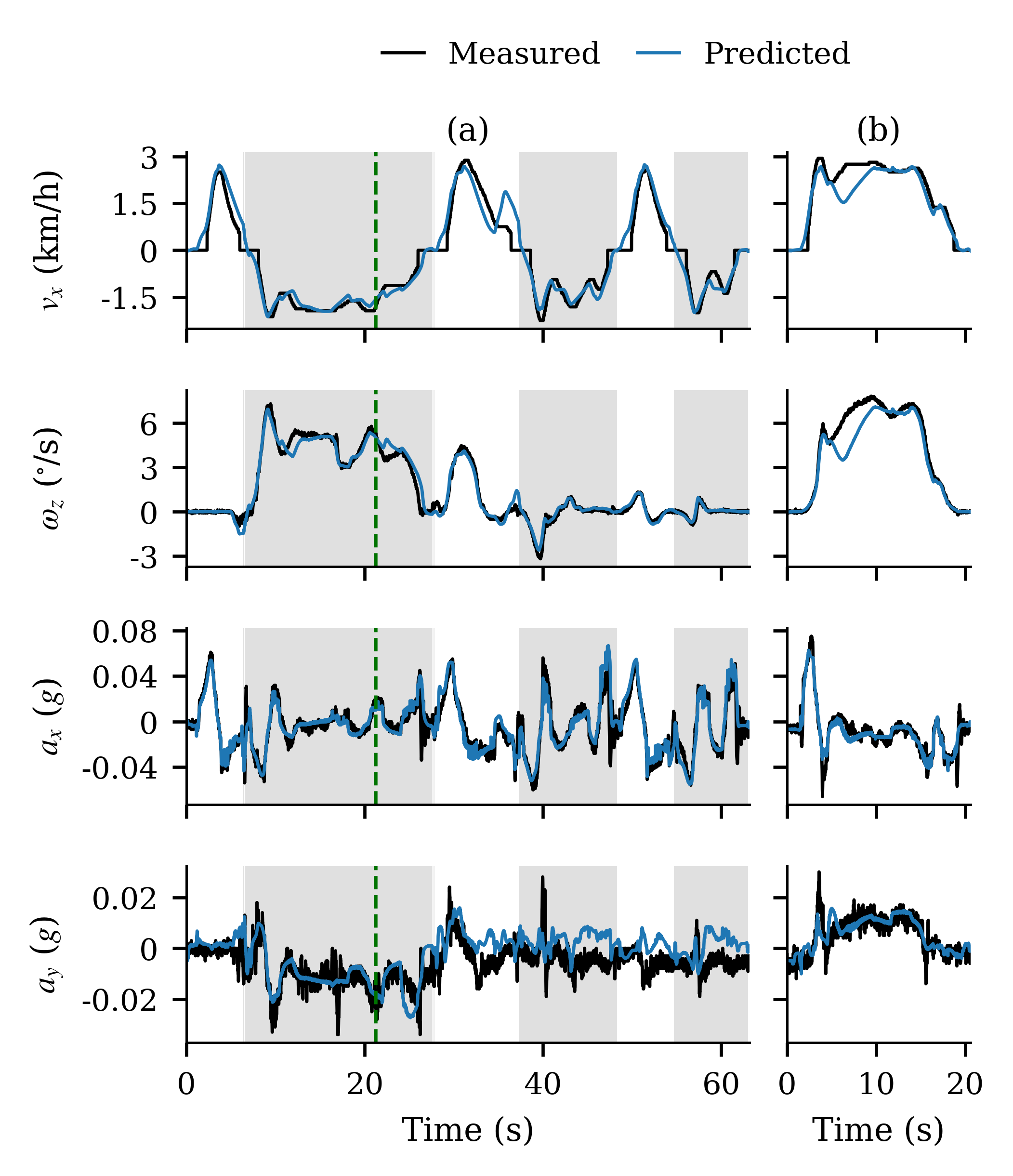}
\caption{Open-loop rollout on two held-out validation maneuvers of the
canonical fold: (a) the multi-segment park-in and (b) the single-segment
park-out, comparing measured and predicted $v_x$, $\omega_z$, $a_x$, and
$a_y$. Rows share axes and the two maneuvers share a common time scale;
shaded bands mark reverse-gear segments, and the green dashed line marks
the training-window horizon.}
\label{fig:val}
\end{figure}

\begin{figure*}[!t]
\centering
\includegraphics[width=\textwidth]{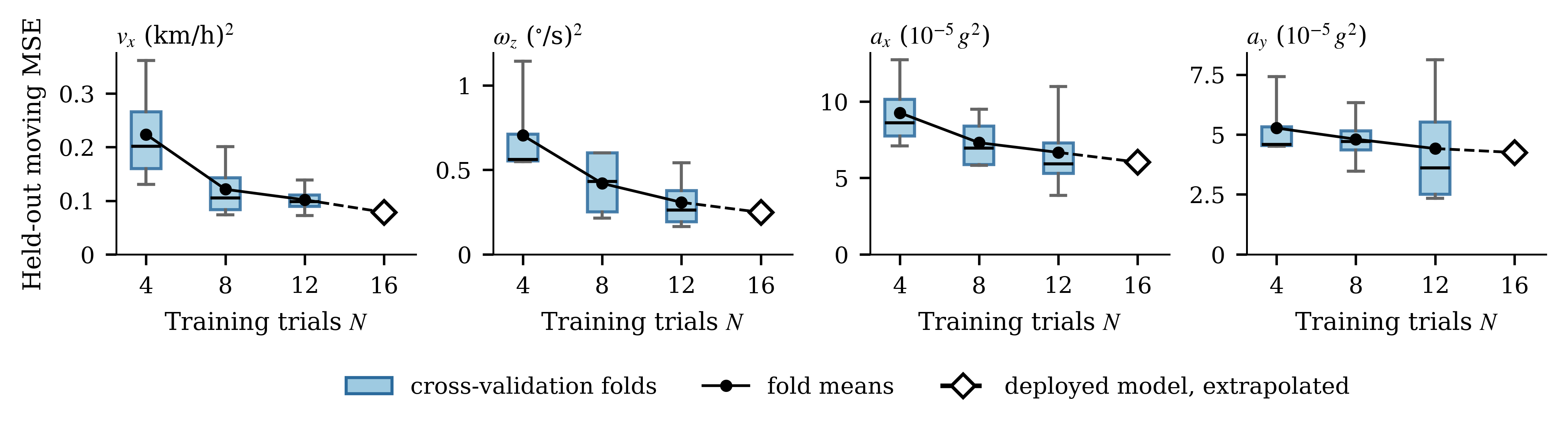}
\caption{Generalization versus training-set size under
leave-$r$-repeats-out cross-validation. Boxes span the four folds at each
training-set size, and the solid line connects the fold means; the
diamond marks the all-16 deployment model, whose held-out error is
extrapolated by a power-law fit of the fold means because no held-out
trial remains, and the dashed segment extends that fit.}
\label{fig:cv}
\end{figure*}

Table~\ref{tab:cv} reports the four-fold cross-validation on the complete
held-out maneuvers in the moving region. The velocity MSE of
$0.108 \pm 0.038$\,(km/h)$^2$ is a root-mean-square error of about
0.33\,km/h on a signal reaching only 3\,km/h. In the open-loop rollouts of
Fig.~\ref{fig:val}, the shaded reverse segments are tracked as closely as
the forward bursts.

\begin{table}[!t]
\caption{Four-Fold Cross-Validation on Held-Out Maneuvers}
\label{tab:cv}
\centering\footnotesize
\begin{tabular}{lccc}
\toprule
Fold & $v_x$ [(km/h)$^2$] & Rev. $v_x$ [(km/h)$^2$] & $\omega_z$ [($^{\circ}$/s)$^2$] \\
\midrule
0 & 0.057 & 0.051 & 0.162 \\
1 & 0.095 & 0.040 & 0.201 \\
2 & 0.121 & 0.065 & 0.575 \\
3 & 0.160 & 0.072 & 0.439 \\
\midrule
mean $\pm$ std & $0.108 \pm 0.038$ & $0.057 \pm 0.012$ & $0.344 \pm 0.170$ \\
\bottomrule
\end{tabular}
\end{table}

The spread across folds has an identifiable origin. The one-step
prediction errors are essentially identical across folds, so the spread
emerges only under multi-step rollout, where occasional transient
excursions at velocity transitions compound. Retraining the fold-1 model
under three further random seeds on the same split yields
$0.089 \pm 0.019$\,(km/h)$^2$, a within-split seed spread of the same
order as the across-fold spread. The cross-validation variance therefore
reflects training stochasticity compounding through the open-loop
rollout, not differences in maneuver difficulty between folds.

Figure~\ref{fig:cv} places these results on the learning curve from the
leave-$r$-repeats-out grid: under the matched 3000-epoch protocol the
velocity MSE falls monotonically from 4 to 12 training trials and
is still descending at the data boundary. The model is therefore
data-limited rather than capacity-limited, so further repetitions would
improve accuracy directly, while the achieved level already suffices for
the closed-loop application of Section~\ref{sec:results:closedloop}.
Long-horizon behavior is likewise bounded: over complete maneuvers up to
three times the training window the full-trajectory errors reach
0.172\,(km/h)$^2$ and 0.363\,($^{\circ}$/s)$^2$ without divergence, and the
beyond-horizon segments are predicted no worse than those within the
training window.

\subsection{The Physics-Informed Loss Replaces the Inference Limiter}
\label{sec:results:limiter}
The parent model~\cite{song2024pynssm} enforces physical plausibility
through a hard state limiter applied at inference, deliberately keeping
such clamping out of training; the physics-informed formulation of
Section~\ref{sec:model:pinn} inverts this philosophy. On the same trained
model over the held-out maneuvers, adding the gear-conditioned kinematic
limiter at inference changes the moving-region velocity error by under
3\%, from 0.160 to 0.156\,(km/h)$^2$, and leaves the reverse-velocity and
yaw-rate errors unchanged, all within the seed-noise floor, because the
training losses \eqref{eq:pinnv}--\eqref{eq:pinnyaw} have already confined
the dynamics to the admissible region. The limiter is therefore redundant,
and removing it spares a per-step branch in the real-time loop. Its yaw
envelope would in any case demand care: it must clamp the
reconstructed yaw rate of \eqref{eq:residual}, and clamping the
residual channel directly pins it to the kinematic manifold, so the
rollout diverges.

\subsection{Design Ablations}
\label{sec:results:ablations}

Every major design decision was resolved by a controlled comparison under
the evaluation protocol of Section~\ref{sec:setup:training};
Table~\ref{tab:ablations} consolidates the principal tuning axes, all
evaluated on complete held-out maneuvers in the moving region.

\begin{table}[!t]
\caption{Design Ablations on Held-Out Maneuvers}
\label{tab:ablations}
\centering\footnotesize\setlength{\tabcolsep}{3.4pt}
\begin{tabular}{llcc}
\toprule
Axis & Option & $v_x$ [(km/h)$^2$] & $\omega_z$ [($^{\circ}$/s)$^2$] \\
\midrule
\multirow{3}{*}{Normalization}
 & clipped min--max$^{\dagger}$ & 0.202 & 0.691 \\
 & plain min--max & \textbf{0.201} & 0.716 \\
 & z-score & 0.243 & \textbf{0.618} \\
\midrule
\multirow{3}{*}{Loss (at its best LR)}
 & MAE$^{\dagger}$ & \textbf{0.163} & \textbf{0.657} \\
 & MSE & 0.173 & 0.701 \\
 & log-cosh & 0.248 & 0.773 \\
\midrule
\multirow{2}{*}{LR schedule}
 & cosine$^{\dagger}$ & \textbf{0.196} & 0.699 \\
 & const. & 0.202 & \textbf{0.691} \\
\midrule
\multirow{3}{*}{Training window $L$}
 & 2121 (no padding)$^{\dagger}$ & \textbf{0.079} & \textbf{0.280} \\
 & 3435 (10\% padding) & 0.103 & 0.327 \\
 & 6390 (39\% padding) & 0.092 & 0.319 \\
\bottomrule
\end{tabular}
\par\vspace{2pt}
\parbox{\linewidth}{\footnotesize Each axis varies one factor with the others held at a stage-specific baseline, so values compare only within an axis. The best value in each column is in bold; $^{\dagger}$marks the option carried into the deployed model.}
\end{table}

\subsubsection{Robust Normalization and Loss}
Clipping each channel to its 0.5th--99.5th percentiles before min--max
scaling is the deployed normalization, chosen to keep the scaled
signals bounded: it ties plain min--max on velocity and trails only z-score
on yaw, and z-score is rejected for the velocity penalty it pays on the
priority variable (Table~\ref{tab:ablations}). The deployed mean absolute
error is the most accurate loss at its best learning rate and, by a wide
margin, the smoothest to train, its late-training variation under 1\%
against 2.6\% for MSE. The learning rate is the principal instrument for
the velocity--yaw trade, since raising it improves the velocity at the
expense of the yaw; the deployed cosine schedule sits at the
velocity-favoring end at $2 \times 10^{-3}$, its yaw within the seed-noise
floor of the constant schedule. The schedule rows of
Table~\ref{tab:ablations} isolate the schedule shape at a reduced learning
rate.

\subsubsection{Short Training Windows Generalize Best to Long Rollouts}
Although every evaluation rolls out over complete maneuvers of up to 6390
steps, training on the shortest 2121-step windows yields the best
full-trajectory and even beyond-horizon accuracy, at a third of the
per-epoch cost of full-length sequences, which spend 39\% of their
computation on padding. The cause is data diversity: the shortest window
tiles the 12 training trials into 29 distinct training sequences whose
gradients sample every maneuver phase, whereas the longer windows reduce
the corpus to 12 to 18 padded sequences that generalize worse.
Backpropagating through a longer horizon does not substitute for gradient
diversity.

\subsubsection{The Physics-Residual Readout Sharpens the Velocity}
Reading the yaw rate as a residual on the kinematic-bicycle prior of
\eqref{eq:bicycle}, rather than predicting it directly, lowers the
moving-region velocity error by 27\% and the yaw-rate error by 14\% against
an otherwise identical six-layer network. That the velocity gains
more, though the readout acts on the yaw, follows from the coupling of
\eqref{eq:bicycle}: anchoring the yaw to its kinematic value spares the
network from reproducing the motion the prior already explains and feeds
the sharper yaw back into the velocity rollout through the shared state.
The readout is retained in every deployed model.

\subsubsection{Integration Order and Capacity Are Null Axes}
Replacing the forward-Euler integration in \eqref{eq:nss:state} by
midpoint or fourth-order Runge--Kutta changes the validation metrics by
under 1\% while costing 1.7 and 3.1 times the training wall clock, so the
cheapest scheme is the correct engineering choice. Capacity behaves analogously: the selected six-layer state network
outperformed both smaller variants and modernized alternatives, namely
augmented latent states and residual connections with layer
normalization, which overfit the 16-trial corpus.

\subsection{Actuator Submodel Selection}
\label{sec:results:actuators}

\begin{figure}[!t]
\centering
\includegraphics[width=\columnwidth]{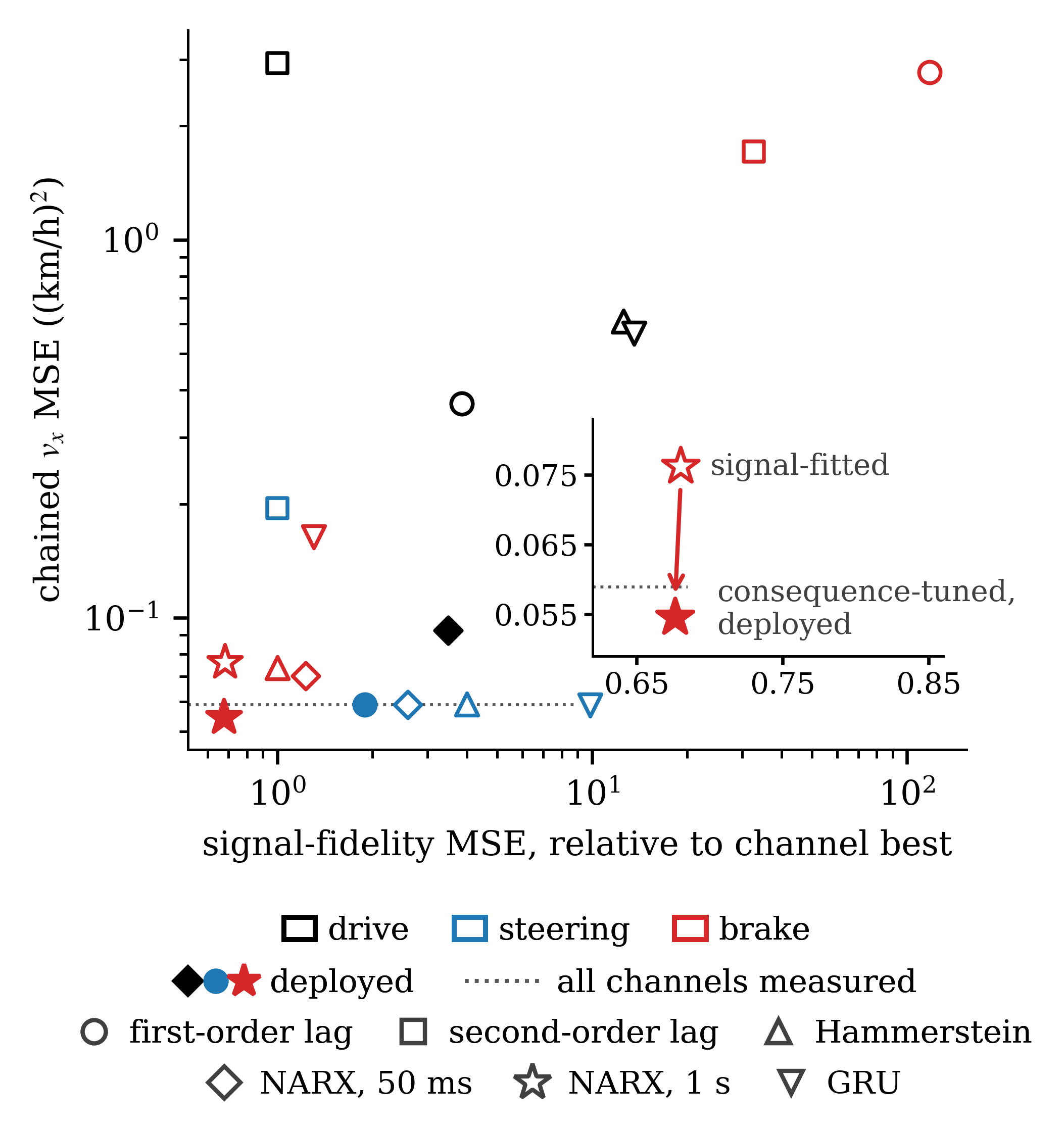}
\caption{Signal fidelity versus end-to-end closed-loop value for the
command-only candidates of the three deployed channels; each point
substitutes one candidate into the chain with all other channels
measured. The two stars are
the 1\,s NARX brake before and after consequence-level co-tuning, and the
inset magnifies its crossing of the reference at unchanged signal
fidelity.}
\label{fig:fidelity}
\end{figure}

Actuator selection on the deployment model is evaluated at two closed-loop
levels, both scored by the full-trajectory velocity MSE against the
all-measured reference: a
per-channel level that substitutes each candidate alone with all other
channels measured (Fig.~\ref{fig:fidelity}), and an assembly level that
activates all submodels at once, summarized as a fidelity ladder in
Table~\ref{tab:actuators}.

\begin{table}[!t]
\caption{Assembled-Chain Fidelity Ladder}
\label{tab:actuators}
\centering\footnotesize\setlength{\tabcolsep}{4pt}
\begin{tabular}{lc}
\toprule
Configuration & $v_x$ [(km/h)$^2$] \\
\midrule
All channels measured (reference) & 0.059 \\
\midrule
Assembled chain, signal-fitted 1\,s NARX brake & 0.069 \\
Assembled chain, consequence-tuned 1\,s NARX brake & 0.055\rlap{$^{\dagger}$} \\
\midrule
Monolithic NSS on commands, no submodels & 0.130 \\
Static affine actuator maps, no dynamics & 0.375 \\
\bottomrule
\end{tabular}
\par\vspace{2pt}
\parbox{\linewidth}{\footnotesize $^{\dagger}$Deployed configuration.}
\end{table}

\subsubsection{Signal Fidelity Does Not Predict Closed-Loop Value}
The Phase-A fidelity screen and the end-to-end criterion disagree sharply.
Across the 15 channel--architecture pairs of the three deployed channels,
candidates at channel-best signal fidelity span chained velocity errors
from near the all-measured reference to a 50-fold degradation, and the two
criteria correlate at only $+0.07$ (Fig.~\ref{fig:fidelity}): the drive
channel's highest-fidelity fit destabilizes the chain, while a long-memory
brake fitted to the recorded signal, though it halves the hold-region
pressure error, degrades the chain because its learned hold transients fall
where pressure errors carry velocity consequences. Submodels for
closed-loop simulation must therefore be selected in the closed loop, and
fidelity gains on autonomously actuated components warrant suspicion.

\subsubsection{State Dependence Differs Across Channels}
Whether actuator dynamics depend on the vehicle state is answered per
channel, and the answer differs. The drive channel benefits
from a uniform state interface, cutting its per-channel error by a third
and reflecting a torque response that varies with speed and gear; the brake is
best modeled from commands alone, every state-aware variant losing in the
assembled chain; and the steering channel is dynamically decoupled at
parking speeds, so its one-parameter lag is chosen for simplicity.
Closed-loop fine-tuning narrows a state-aware submodel's
identification-to-deployment gap but neither reverses an input-set
disadvantage nor transfers across assemblies, so the deployed drive
submodel is the identification-time one.

\subsubsection{Consequence-Level Co-Tuning and the Value of Structure}
With every submodel active, the assembled chain is the governing
criterion, stricter than signal fidelity or per-channel substitution
because submodel errors interact through the shared vehicle state; its gap
from the all-measured reference in Table~\ref{tab:actuators} is the cost of
replacing every measured signal by its submodel. The envelope stress study
of Section~\ref{sec:actuators:limits} shows the input-domain clamp is the
active regulator under out-of-domain commands, the output clamp a
zero-cost backstop.

The long-memory failure above is a property of the training objective, not
of memory itself. Fine-tuning the 1\,s NARX end-to-end, by backpropagating
the blind-zone-masked velocity loss through a differentiable replica of the
frozen chain into the brake weights alone, reverses the verdict. The
co-tuned submodel reaches 0.055\,(km/h)$^2$ full-trajectory and 0.043 in
the moving region, against 0.059 and 0.049 for the 50\,ms NARX under
identical runtime, retaining the halved hold-region pressure error that
signal-level training could not convert into closed-loop value
(Fig.~\ref{fig:fidelity}). The consequence loss weighs pressure errors by
their effect on the vehicle state, a learned soft masking that outperforms
an explicit hard mask. The full-trajectory figure dips below the
all-measured reference, a co-adaptation to the vehicle model's biases
rather than accuracy beyond measurement, while the masked moving-region
error stays above its 0.037 reference; signal-level fine-tuning never
reversed an input-set disadvantage, whereas consequence-level
co-optimization reversed an architecture-level one.

Two baselines without submodels bound the value of the chain's structure
in Table~\ref{tab:actuators}. Static affine calibrations inflate the
velocity error to nearly 7 times the deployed chain, and retraining the NSS on
commanded inputs, absorbing the actuators with no submodels, reaches 2.4
times: the two-dimensional vehicle state carries no actuator memory, so a
command-driven network has nowhere to store the torque build-up, hold
pressure, and release latency that the submodels hold explicitly in their
command histories. The modular chain is the accuracy-optimal arrangement of
the same information.

The accuracy margin, however, is not the principal argument for the
decomposition. The subsystems of an automated-parking stack evolve on
independent schedules, and the modular chain absorbs a brake, steering-rack,
or power-steering change by re-identifying the affected submodel alone,
reusing the NSS and the other channels unchanged, whereas a monolithic
model entangles every subsystem and forces a complete re-collection and
retraining. Modularity thus converts vehicle-level retraining into
component-level recalibration, which matters more than accuracy for
sustained closed-loop use.

\subsection{Standardized Fidelity of the Deployed Chain}

\begin{figure}[!t]
\centering
\includegraphics[width=\columnwidth]{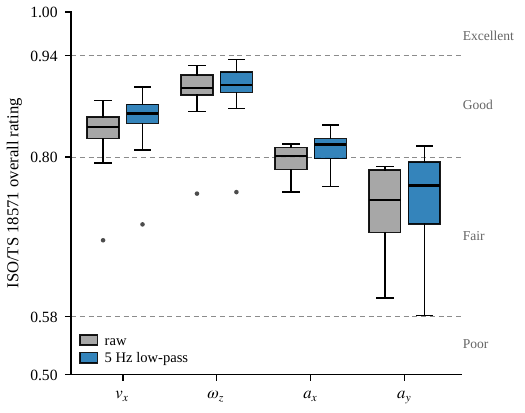}
\caption{Objective rating of the deployed command-to-vehicle chain against the
measured field tests under ISO/TS~18571, per channel over the 16
maneuvers. Each box spans the 16 maneuver ratings; every maneuver is scored
both raw and after an identical 5\,Hz zero-phase low-pass on the measured and
predicted signals. Dashed lines mark the ISO/TS~18571 grade thresholds.}
\label{fig:iso}
\end{figure}

The mean-square errors of the fidelity ladder quantify accuracy in
physical units but give no standardized measure of how closely the
deployed plant reproduces the instrumented vehicle. The assembled
command-to-vehicle chain is therefore additionally assessed under
ISO/TS~18571~\cite{iso18571}, an objective-rating metric that scores the
correlation between a measured and a simulated time history, combining
corridor, phase, magnitude, and slope sub-ratings into a single
zero-to-one rating graded Excellent, Good, Fair, or Poor. From the
recorded command stream the chain predicts four channels, $v_x$,
$\omega_z$, $a_x$, and $a_y$, each rated against its measured counterpart
over the complete maneuver across all 16 field tests. The rating is
reported in two forms: the measured acceleration and yaw-rate signals
carry wheel-speed and inertial-sensor noise that the smooth model neither
reproduces nor should, and because the slope sub-rating is sensitive to
it, this noise depresses the rating without reflecting model error.
ISO/TS~18571 accordingly prescribes identical conditioning of both
signals~\cite{iso18571}, so a 5\,Hz zero-phase low-pass is applied to measured and predicted alike,
with the raw rating retained alongside for transparency.

As Fig.~\ref{fig:iso} shows, the deployed chain attains Good
ratings on the two state variables, $v_x$ at 0.83 and $\omega_z$ at 0.89,
and ranges from the Good--Fair boundary at the longitudinal acceleration
$a_x$, 0.80, to Fair at the lateral acceleration $a_y$, 0.72,
averaging 0.81 raw and 0.82 filtered across the four channels. The phase
sub-rating is near unity throughout, while the slope sub-rating is the common
limiter, consistent with the residual sensor noise that the 5\,Hz
conditioning partly relieves. The low $v_x$ and $\omega_z$ outliers come
from the single maneuver with the largest open-loop velocity excursion at
its motion-direction transitions, yet its ratings remain within the
Fair band. Parking confines the acceleration channels to
small-amplitude, low signal-to-noise dynamics, most acutely in the lateral
axis, so their lower ratings relative to the state variables are expected,
and that they remain Fair indicates the chain captures their
essential character. These ratings characterize the reconstruction
fidelity of the deployed all-16-maneuver model; held-out generalization is
reported separately by Table~\ref{tab:cv}.

\subsection{Closed-Loop Evaluation in the Parking Simulator}
\begin{figure}[!t]
\centering
\subfloat[]{\includegraphics[width=0.49\columnwidth]{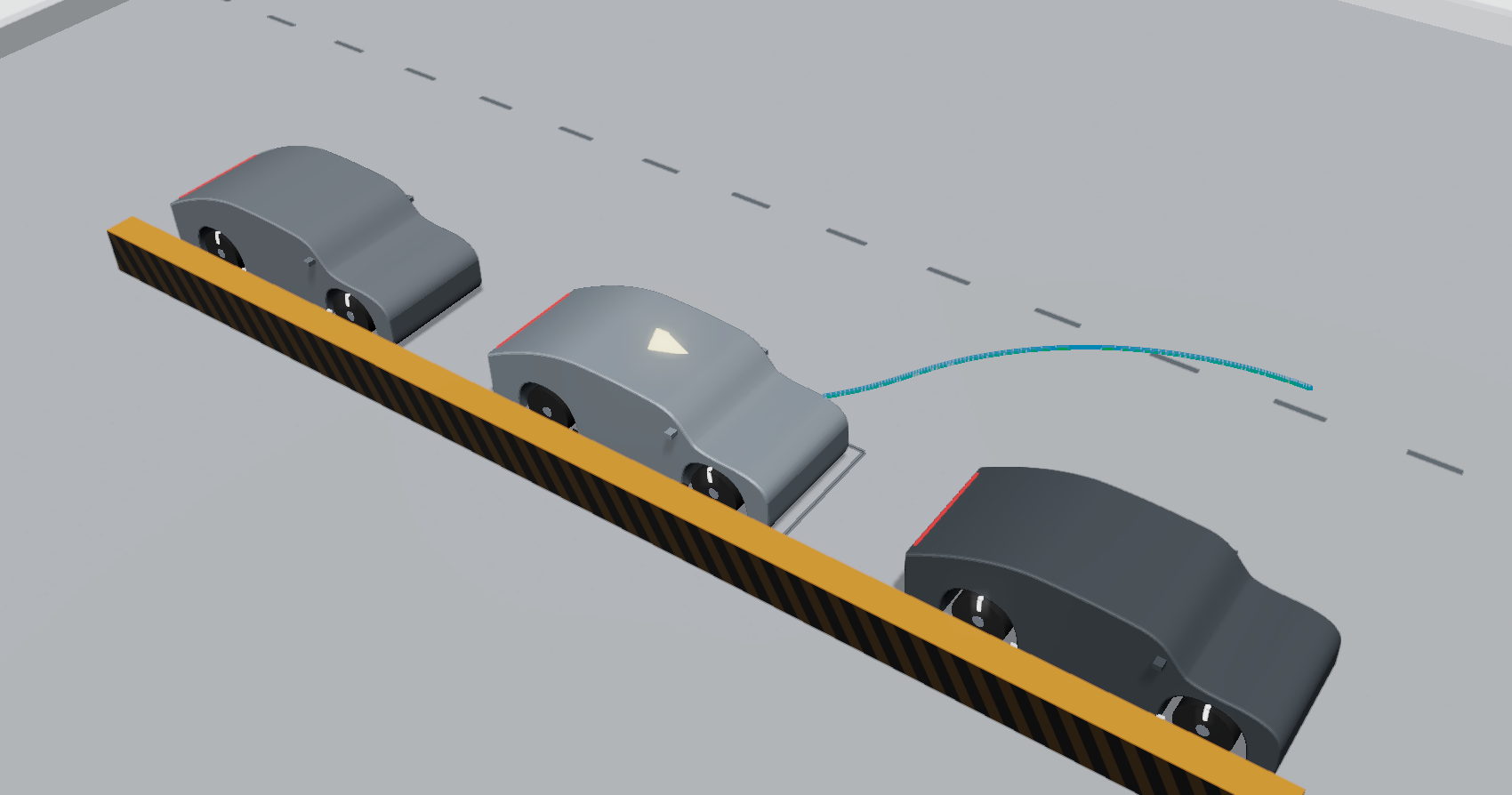}}
\hfil
\subfloat[]{\includegraphics[width=0.49\columnwidth]{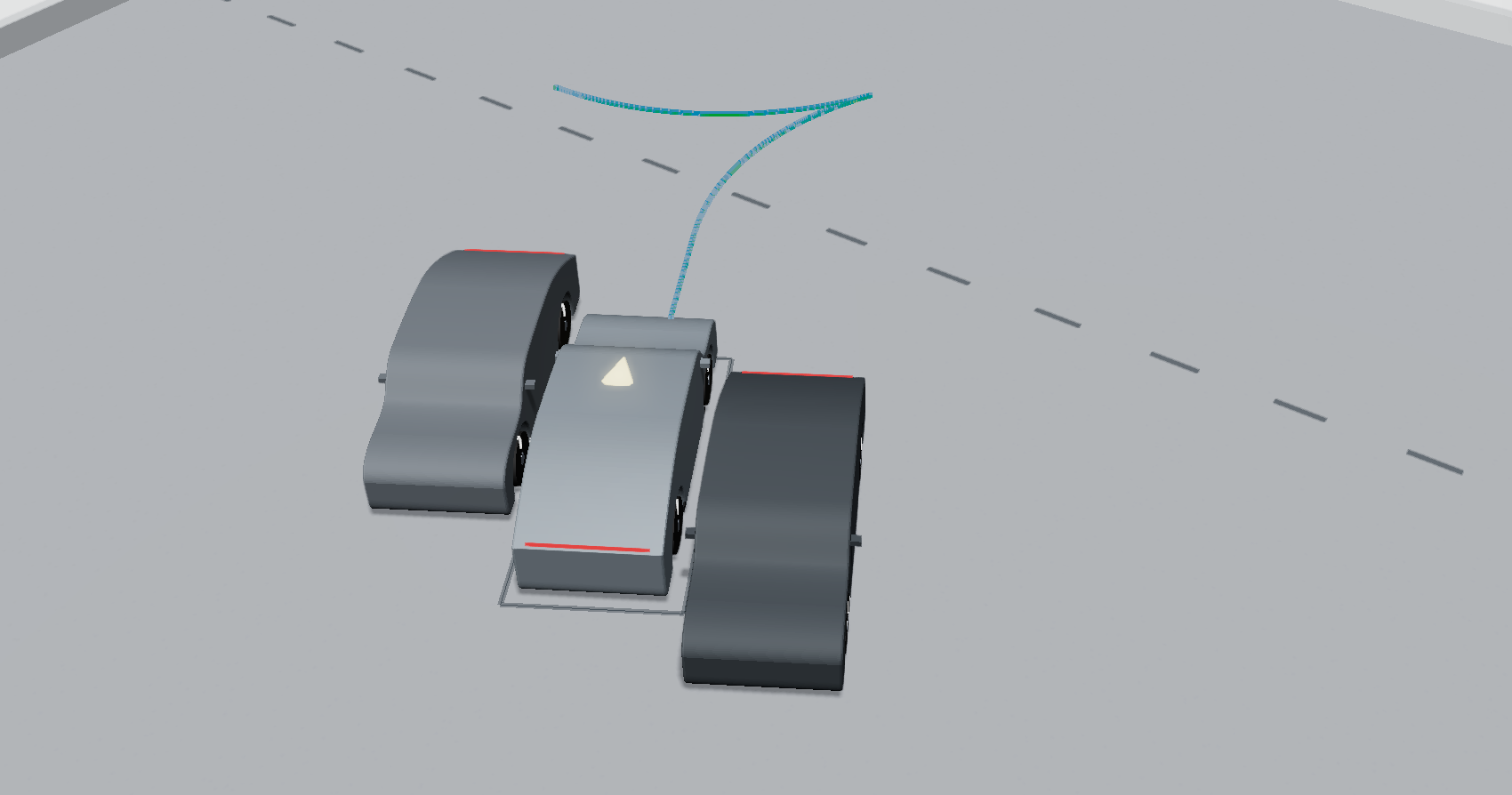}}
\caption{Closed-loop execution of planned maneuvers by the NSS plant in
the parking simulator: (a) the two-cusp parallel maneuver and (b) the
one-cusp angled back-in maneuver. The planned path is drawn in green and
the driven path in cyan; the two nearly coincide.}
\label{fig:sim}
\end{figure}

\label{sec:results:closedloop}

The test of a vehicle model intended for simulation is whether a
complete planning and control stack can park the car through it.
Table~\ref{tab:closedloop} compares executed maneuvers over the
deterministic 36-cell scenario grid, spanning three bay types, two
approach directions, three difficulty tiers, and the lane barrier present
or absent, under the ideal kinematic bicycle and the learned plant chain.
The plan is identical in both, so the comparison isolates the cost of
realistic dynamics.

\begin{table}[!t]
\caption{Closed-Loop Execution Over the Scenario and Robustness Grids}
\label{tab:closedloop}
\centering\footnotesize\setlength{\tabcolsep}{4pt}
\begin{tabular}{lrr}
\toprule
 & \multicolumn{1}{c}{Kinematic} & \multicolumn{1}{c}{NSS$^{\dagger}$} \\
\midrule
Completed maneuvers & 31/31\phantom{.0} & 31/31\phantom{.0} \\
Goal error, mean [cm] & 1.0 & 4.5 \\
Goal error, median [cm] & 0.8 & 3.5 \\
Goal error, max [cm] & 2.3 & 10.0 \\
Driven clearance, mean [cm] & 33.0 & 27.6 \\
Cells under the 20\,cm sensor floor & 0\phantom{.0} & 10\phantom{.0} \\
\midrule
\multicolumn{3}{l}{Held-out axes, NSS plant; goal error
mean/med/max [cm]} \\
Obstacle skew $\pm 5^\circ$ (26 cells) & \multicolumn{2}{c}{\phantom{0}5.1 / 4.9 / \phantom{0}10.1} \\
Start jitter $\pm 0.25$\,m, $\pm 3^\circ$ (52) & \multicolumn{2}{c}{\phantom{0}7.6 / 4.9 / 114.8} \\
Tighter slots, $-7.5$\,cm (27) & \multicolumn{2}{c}{\phantom{0}9.0 / 4.5 / 105.3} \\
Looser slots, $+7.5$\,cm (34) & \multicolumn{2}{c}{18.4 / 4.8 / 463.9} \\
\bottomrule
\end{tabular}
\par\vspace{2pt}
\parbox{\linewidth}{\footnotesize $^{\dagger}$Deployed execution plant.}
\end{table}

\begin{figure}[!t]
\centering
\includegraphics[width=\columnwidth]{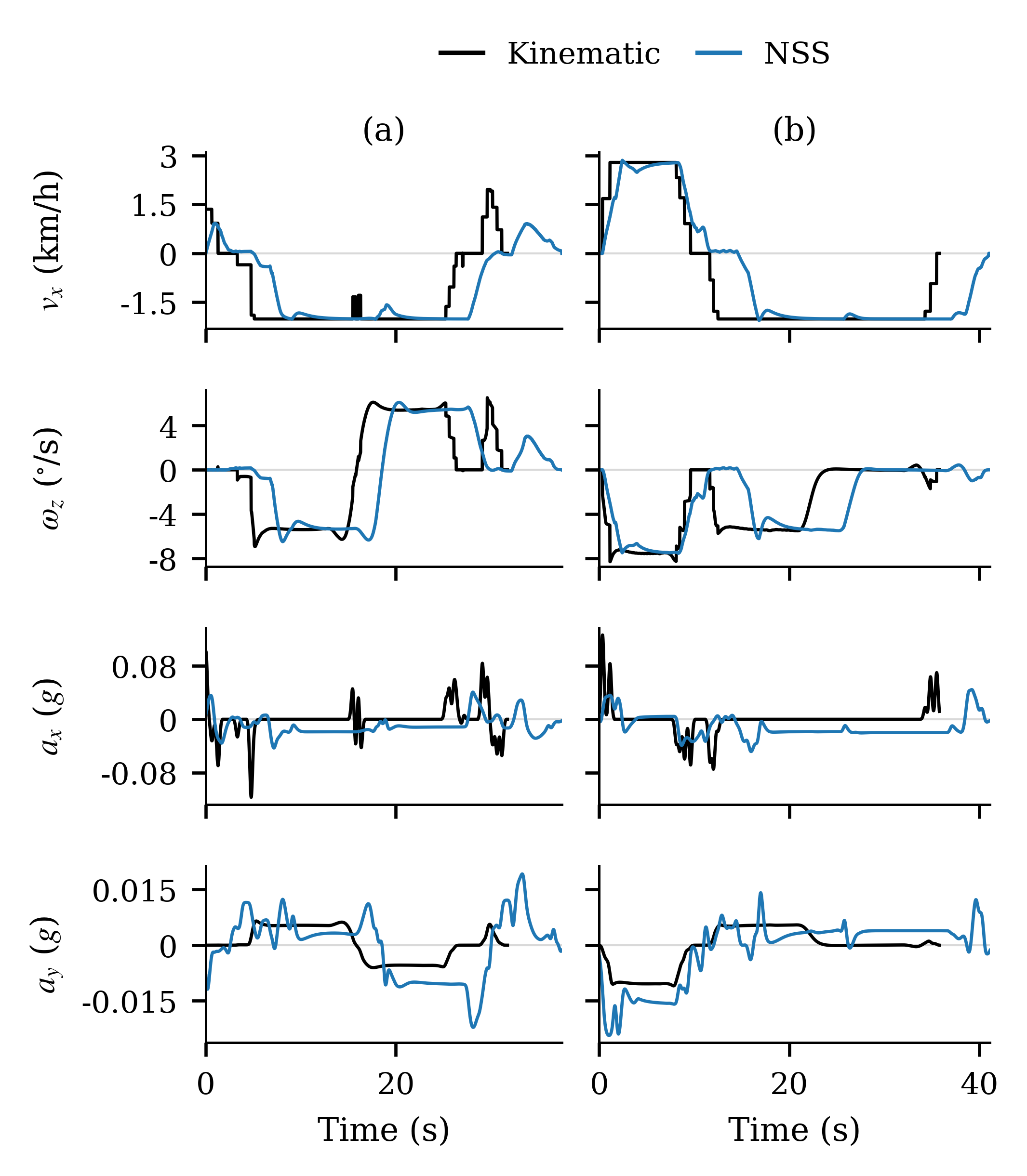}
\caption{Vehicle-state trajectories under the ideal kinematic plant and the
NSS plant for the two maneuvers of Fig.~\ref{fig:sim}: (a) the two-cusp
parallel maneuver and (b) the one-cusp angled back-in. An identical plan and
predictive controller drive both plants, so the difference isolates the plant
dynamics. The two columns share a common time scale, and $a_x$ and
$a_y$ are shown with a short smoothing window.}
\label{fig:simcompare}
\end{figure}

Under the learned dynamics every plannable maneuver completes, parking to
a median goal error of 3.5\,cm against the kinematic plant's idealized
0.8\,cm, centimeter-level accuracy through real actuator lag, drivetrain
creep, and brake-hold, executed in real time in the browser
(Fig.~\ref{fig:sim}). This required the longitudinal control of
Section~\ref{sec:setup:sim}; without it the open-loop replay of the
planned speed profile diverges, exactly as a real vehicle drifts from a
clock-indexed plan. Under the same plan and controller, the learned
dynamics stretch each maneuver some 15--18\% beyond the kinematic execution
(Fig.~\ref{fig:simcompare}), yet the controller holds both to the same
parked pose, while 10 cells in the tightest tiers finish below the
20\,cm ultrasonic clearance floor the kinematic execution maintains by
construction.

This clearance erosion is a controller-level trade, not a model
deficiency. The predictive tracker cuts each corner slightly inside the
plan, holding landing accuracy under 5\,cm while shrinking the swept
clearance, and disabling the feedback reverses the balance. The trade is
fundamental to executing a kinematic plan on realistic dynamics; its
production resolutions, replanning at junctions or clearance-aware
tracking, lie beyond this paper's scope, and the grid is reported with the
trade visible rather than tuned away.

The lower block of Table~\ref{tab:closedloop} probes overfitting of the
interface tuning: the gains tuned on the 12-cell core are evaluated on a
252-cell extended grid of obstacle, start-pose, and slot-width
perturbations never seen during tuning. The typical held-out cell parks at
core-grid accuracy, with no cell changing solvability between the two
plants; degradation is a tail phenomenon, five of the 139 plannable cells
exceeding 20\,cm where long brake-hold phases meet far-off-nominal
geometry, the worst a loosened-slot topology that drives the command
interface into a livelock far from the goal. These tails proved sensitive
to the brake submodel: an intermediate deployment with a less accurate
brake doubled the tail count under identical gains, identifying the plant,
not the controller tuning, as the binding factor for held-out robustness.

Beyond the gross motion, the learned plant carries the measured signal
characteristics of the instrumented vehicle into the virtual phase, as the
acceleration channels of Fig.~\ref{fig:simcompare} show. Where the
kinematic plant emits only the accelerations its idealized motion implies,
the NSS plant reproduces them as the on-board inertial sensor reports them,
with the offset and attitude coupling that sensor mounting and load
transfer introduce and the point-mass idealization omits. A parking controller consuming
these signals, for ride-comfort shaping, grade or load compensation, or
stop detection, can thus be tuned against realistic measurements before
the target vehicle is instrumented.

\section{Conclusion}
\label{sec:conclusion}
This study develops a physics-informed neural state-space model tailored
for the parking-regime dynamics of a production battery-electric sedan,
identified entirely from field-test maneuvers and carried through to a
real-time closed-loop deployment. By favoring a data-driven formulation
grounded in physics over a fully parameterized description, this work
captures the actuator lag, drivetrain creep, and direction reversals that
the kinematic idealization omits at parking speeds. The gear-conditioned
velocity constraint and the physics-residual yaw readout divide the labor
between physics and learning, so that imposing the physical constraints
during training makes the customary inference-time state limiter redundant.
The accompanying actuator study shows that a submodel must be judged inside
the deployed assembly rather than by its signal fidelity, and that tuning
the brake on its velocity consequence reverses the verdict reached against
the long-memory architectures at the signal level.

The model generalizes to held-out maneuvers in fully open-loop
simulation, and, despite being identified from only 16 field tests,
the assembled command-to-vehicle chain earns Good ratings on the vehicle
states under the ISO/TS 18571 objective rating metric. Embedded as the
real-time plant of an interactive simulator, it enables a
production-representative planning stack to park the vehicle through the
learned dynamics. Because the model is identified entirely from field-test
logs of the target vehicle rather than from a parameterized dynamics
description, a simulator built around it can pre-calibrate an
automated-parking planning and control stack in the virtual development
phase without the manufacturer's proprietary chassis and actuator
parameters. Future efforts will focus on broadening the single-vehicle
data scale and extending the model to a wider range of road surface
conditions, such as varying slopes and road adhesion coefficients. A
further step is to unify this parking-regime model with the
forward-driving dynamics of its precursor~\cite{song2024pynssm} into a
single neural state-space model spanning the full speed envelope.

\section*{Acknowledgment}
The authors would like to express their gratitude to
Mrs.~Grace (Shenghan) Gao for her guidance and advice on the
writing and grammatical aspects of this paper.

\end{document}